\documentclass[conference]{IEEEtran}
\IEEEoverridecommandlockouts
\usepackage{cite}
\usepackage{amsmath,amssymb,amsfonts}
\usepackage{algorithmic}
\usepackage{graphicx}
\usepackage{textcomp}
\usepackage{xcolor}
\usepackage{color, colortbl}
\usepackage{multirow, multicol}
\usepackage{adjustbox}
\usepackage{bm}

\def\BibTeX{{\rm B\kern-.05em{\sc i\kern-.025em b}\kern-.08em
    T\kern-.1667em\lower.7ex\hbox{E}\kern-.125emX}}

\usepackage{fancyhdr}
\begin{document}

\title{Closing the Affective Loop via Experience-Driven Reinforcement Learning Designers\\
\thanks{This project has received funding from the Malta Council for Science and Technology through the SINO-MALTA Fund 2022, Project OPtiMaL. Diogo Branco was supported by Fundação para a Ciência e Tecnologia (FCT) under the PhD grant 2021.05646.BD and by the Project ``Consortium Advanced Computing - HPC, HPDA, AI \& HPV'' (2021-1-PT01-KA131-HED-000008876) under the Erasmus+ Program, Education and Training 2021 - Higher Education - Action KA131 - Mobility for Learning Purposes. }
}

\author{
    \IEEEauthorblockN{Matthew Barthet\textsuperscript{1}, 
                      Diogo Branco\textsuperscript{2}, 
                      Roberto Gallotta\textsuperscript{1}, 
                      Ahmed Khalifa\textsuperscript{1}, 
                      Georgios N. Yannakakis\textsuperscript{1}}
    \IEEEauthorblockA{\textsuperscript{1}Institute of Digital Games, University of Malta, Msida, Malta \\
                      Email: \{matthew.barthet, roberto.gallotta, ahmed.khalifa, georgios.yannakakis\}@um.edu.mt}
    \IEEEauthorblockA{\textsuperscript{2}Faculty of Exact Sciences and Engineering, University of Madeira, Madeira, Portugal \\
                      Email: diogo.branco@arditi.pt}
}

\maketitle
\thispagestyle{fancy}

\begin{abstract}
Autonomously tailoring content to a set of predetermined affective patterns has long been considered the holy grail of affect-aware human-computer interaction at large. The experience-driven procedural content generation framework realises this vision by searching for content that elicits a certain experience pattern to a user. In this paper, we propose a novel reinforcement learning (RL) framework for generating affect-tailored content, and we test it in the domain of racing games. Specifically, the experience-driven RL (EDRL) framework is given a target arousal trace, and it then generates a racetrack that elicits the desired affective responses for a particular type of player. EDRL leverages a reward function that assesses the affective pattern of any generated racetrack from a corpus of arousal traces. Our findings suggest that EDRL can accurately generate affect-driven racing game levels according to a designer's style and outperforms search-based methods for personalised content generation. The method is not only directly applicable to game content generation tasks but also employable broadly to any domain that uses content for affective adaptation.

\end{abstract}


\section{Introduction}

One of the most challenging tasks within affective computing (AC) \cite{picard2000affective} is to effectively leverage affect models to autonomously generate new contexts that are, in turn, capable of eliciting a desired emotional response \cite{Experience-Driven-PCG}. In other words, the challenge for AC is to successfully enable an affect-aware closed-loop adaptive system, largely known as the \emph{affective loop} \cite{hook2008affective,yannakakis2023affective}. What makes the design of such affect-aware adaptive interactions very difficult is the unpredictability and subjectivity of human users, both behaviourally and emotively. 

\begin{figure}[tb]
    \centering
    \includegraphics[width=0.8\columnwidth]{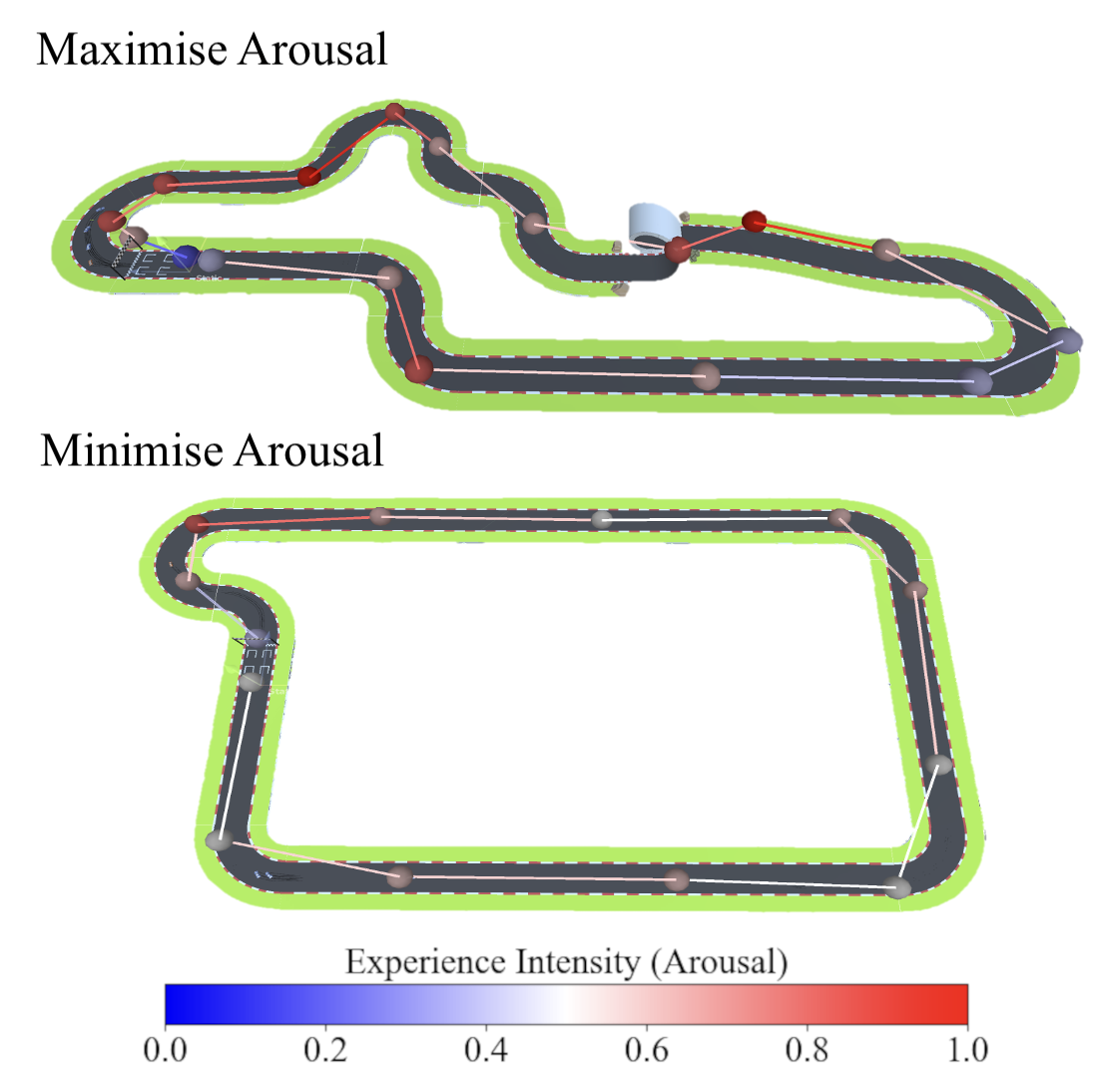}    
    \caption{Two examples of maximally and minimally arousing tracks generated by EDRL for the \emph{Solid Rally} racing game. Top image: EDRL offers content that constantly elicits higher levels of arousal, such as loops and sequences of alternating turns. Bottom image: EDRL generates simple and straight tracks, leaving limited room for highly arousing gameplay. Blue overlay elements denote regions with a strong decrease in arousal, while red denotes regions of a strong increase in arousal, and white denotes neutral arousal changes. }
    \label{fig:solid_rally}
    \vspace{-10pt}
\end{figure}

Motivated by this challenge, we introduce a novel method for autonomously generating content that, when experienced, elicits a desired sequence of emotional responses to a particular user. To test the proposed framework, we focus on the domain of games, as they offer rich forms of human-computer interaction and have proven to be an ideal test bed for AC research in the past \cite{yannakakis2023affective, hudlicka2008affective}. In particular, we leverage human arousal demonstrations from over 100 annotators of a real-time 3D rally driving game from the Arousal video Game AnnotatIoN (AGAIN) corpus \cite{melhart2022arousal} to generate racetracks that elicit a desired arousal trace for a particular player type. To accomplish this, we build upon the experience-driven procedural content generation (PCG) via reinforcement learning (EDRL) framework \cite{EDPCGRL} by using a simulation-based approach to reward the racetracks EDRL generates, as visualised in Fig. \ref{fig:solid_rally}. We compare the performance of the EDRL approach against an experience-driven PCG \cite{EDPCGRL} method that uses genetic search for creating racetracks. Both methods employ the K-Nearest Neighbours (KNN) algorithm \cite{peterson2009k} to generate arousal traces from provided game states encountered during simulations. The result is a generative AI framework that can adapt its generated outcome to multiple player and annotator types, and can be refined over time with more data.

We assess the efficacy of our methods by testing them across three different clusters of human annotators and three indicative target arousal patterns. We both quantitatively compare the ability of the methods to match the target arousal traces, and we conduct an in-depth analysis of the content generated and their associated play traces. Our findings suggest that the introduced EDRL method for continuous affect-driven generation is more efficient and robust than any other method tested. Moreover, findings suggest that certain affective patterns are harder to elicit---via level generation---than others for particular player types. Unsurprisingly, it is easier to generate maximally arousing tracks than minimally arousing tracks for players whose annotated arousal levels are kept low in this game. Inversely, players with strong increases and high arousal values were more difficult to satisfy emotionally when EDRL is tasked to minimize their arousal changes.

\section{Related Work}

In this section, we first cover related studies at the intersection of AC and games in general (Section \ref{sec:affective_computing}) and then specifically AC studies as applied to procedural content generation (Section \ref{sec:pcg}).

\subsection{Affective Computing in Games}
\label{sec:affective_computing}

Applications of AC span a vast array of contexts, user modalities, and domains, including text and natural language \cite{broekens2023fine}, videos \cite{girard2023dynamos}, audio \cite{kim2008emotion}, and games \cite{yannakakis2014emotion,yannakakis2023affective}. Video games present their own rich and unique form of human-computer interaction \cite{barr2007video} in that they allow the user to play an active role during consumption and encompass multiple modalities. Traditionally, models of affect learn to map one or more of these input modalities (e.g., facial input \cite{guo2020real}, pixels \cite{makantasis2021pixels}, game states \cite{melhart2022arousal}) to an affect label reported by a human annotator using supervised learning. Reliably collecting such labels for affect, however, is a significant challenge with a dedicated field of research within AC. Some modern data collection platforms, such as CARMA \cite{girard2014carma} and the PAGAN framework \cite{melhart2019pagan}, enable the collection of such labels in real-time. Recent tools such as RankTrace \cite{lopes2017ranktrace} allow for the collection of unbounded time-continuous signals, which can be used, in turn, for regression or classification tasks, or converted into ordinal labels for preference learning methods \cite{yannakakis2018ordinal}. 

Due to the several challenges involved in the collection of reliable affect labels, an alternative practice in AC is the analysis of an existing affective corpus. Standardized Emotion Elicitation Databases (SEEDs), allow the study of emotion by replicating real life in controlled settings. SEEDs have been solicited through multiple affect elicitors including images \cite{lang1999international,marchewka2014nencki,kurdi2017introducing}, videos \cite{carvalho2012emotional,baveye2015liris,gnacek2024avdos}, and even 3D objects \cite{popic2020database,peeters2018standardized,tromp2020openvirtualobjects}.
In this study, we use the AGAIN dataset \cite{melhart2022arousal}, comprising $1,100$ in-game videos and self-reported annotations of arousal across 124 participants and 9 games. The data consists of in-game telemetry, synchronised with video recordings of the participants' gameplay and time-continuous arousal signals using RankTrace \cite{lopes2017ranktrace}. Models trained on affect labels from the \emph{Solid Rally} racing game of AGAIN have already shown promising results on predicting arousal by solely relying on pixel and in-game audio \cite{makantasis2021pixels} but also training of human-like game playing agents \cite{barthet2022generative}. 

In this paper, we build on earlier studies \cite{barthet2022generative, barthet2022play} and extend the functionality of affect models for the purpose of generating environments that are tailored toward desired affect patterns. We thus attempt to close the affective game loop \cite{yannakakis2023affective} through the game-level generation capacities of the algorithms introduced here.

\subsection{PCG for Affective Computing} \label{sec:pcg}

Since its inception a few decades ago, PCG has evolved to be a hugely influential and critical area of research within the domain of generative media. The initial focus of PCG methods was on adding replayability and novelty to games, such as by generating infinite levels in \textit{Rogue} (Epyx, 1980), which spawned an entire genre of games referred to as \textit{Rogue-likes} \cite{PCG-Survey}. With the recent advent of large language models, the scope for generative systems in games has expanded substantially \cite{gallotta2024large}. Large language models such as GPT have proven themselves capable of both generating text based on emotions \cite{schaaff2023exploring} and processing and recognizing emotions \cite{broekens2023fine}. Whilst affect-driven generation models for other domains remain in its infancy \cite{nie2022review}, such methods have shown promise in domains such as music generation \cite{miyamoto2020music}, facial expression transformation \cite{wu2020cascade}, as well as dialogue generation \cite{colombo2019affect}. 

Within games, affect-based generation has been instantiated by and mostly been explored through the \emph{experience-driven PCG} (EDPCG) framework \cite{Experience-Driven-PCG}. EDPCG aims to generate content that elicits a particular player experience when played. EDPCG primarily takes the form of a search-based PCG method \cite{PCG-Search-Based}, which generates content through algorithmic means such as local search or evolutionary search, allowing for more complex outputs given the right fitness function. Early examples of EDPCG applied in games involve generating personalised levels using some theory-based metric, such as generating racetracks according to \textit{Koster's} \cite{koster2013theory} fun metric \cite{togelius2007towards}, and evolving interesting first-person shooter maps that maximize the duration of close fighting between players \cite{cardamone2011evolving}. Shaker et al. also generated levels for \textit{Super Mario Bros.} (Nintendo, 1985) according to predicted states of the player including engagement, frustration, and challenge \cite{shaker2012evolving}. More recent examples of EDPCG include the generation of video game levels for the \textit{Sonancia} \cite{lopes2015sonancia} system according to the tension experienced by players \cite{lopes2016framing}. 

Recently, EDPCG was combined with PCGRL \cite{khalifa2020pcgrl} to generate game levels for \textit{Super Mario Bros} (Nintendo, 1985) \cite{EDPCGRL} using a reward function which moderates diversity, inspired by \textit{Koster's} principle of \emph{fun} \cite{koster2013theory}. The EDRL framework, in short, has also been extended to generate levels using an episodic soft-actor critic algorithm, allowing it to better tailor itself to individual players \cite{wang2022fun}. To our knowledge, however, EDRL has yet to tackle affect-aware generation using human affect annotations in a continuous manner. This paper introduces the first EDRL agent that relies on time-continuous affect models that, in turn, allow an RL designer to incrementally build content according to a target player type and target affect pattern. Our introduced EDRL agent can either act as a novel-level generator or an affect-aware design assistant.

\section{Experience-Driven Content Generation} \label{sec:algorithm}

\begin{figure}[!tb]
    \centering
    \includegraphics[width =0.9\columnwidth]{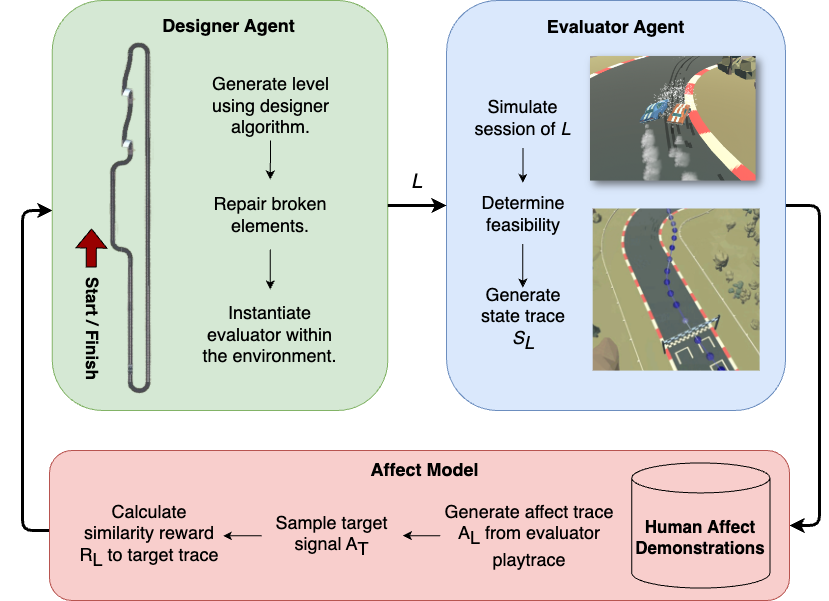}
    \caption{High-level overview of our EDRL framework, with visual examples from our case study detailed in Section \ref{sec:edrl_solid}. \textit{L} refers to the level generated by the designer, $R_L$ refers to the reward function (Eq. \ref{eq:reward}), $S_T$ and $A_T$ are the state and affect traces generated by the evaluator during testing, and $A_T$ is the target affect signal the agent is trying to imitate.}
    \label{fig:algorithm_visual}
\end{figure}

In this paper, we propose a novel approach for generating video game content using a model of human affect demonstrations. The algorithm builds upon the EDRL \cite{EDPCGRL} framework by using a data-driven approach for evaluating generated levels through an evaluator agent combined with an affect model, shown in Fig. \ref{fig:algorithm_visual}. We describe our approach in detail in Section \ref{sec:algorithm} and the different reward functions tested in Section \ref{sec:reward}.

\subsection{Designer}

The designer is responsible for generating the stimuli (racetrack in case of \emph{Solid Rally}) that will be passed to the evaluator agent (hereafter the \textit{evaluator}) during optimisation. Whilst the designer can use any generation method (e.g., PCGML \cite{PCG-ML}, PCGRL \cite{khalifa2020pcgrl}), we test two implementations in this paper: one which takes the form of a search-based generator using a evolutionary RL method (i.e. genetic algorithm), and an RL designer using a variant of the Go-Explore algorithm \cite{ecoffet2021first} called Go-Blend \cite{barthet2022generative} (see Section \ref{sec:edrl_solid}). Regardless of implementation, the output of the designer is a level ($L$) that is assigned a reward based on a function used by the evaluator. Based on this loop of generating new levels followed by feedback from the evaluator's simulation data, the designer should be able to optimise its output over time. Once complete, it should be capable of generating new stimuli that satisfy the target affect trace for a particular player type. 

\subsection{Evaluator}

The role of the evaluator is two-fold. First, the evaluator ensures the feasibility of the generated stimuli (i.e., the generated racetrack) if required during training. If the evaluator identifies a stimulus as infeasible (e.g., an unplayable level featuring a track intersecting itself), the individual is penalised to discourage the designer from re-using this design. Second, the evaluator is responsible for generating the state ($S_L$) and affect ($A_L$) traces for the given stimulus. In this paper, we test an evaluator that generates traces by simulating the playback of the stimuli using an AI agent. 

One more critical design decision for the evaluator is its behaviour during simulation. Due to the data-driven nature of this framework, the affect model and behaviour evaluator can be tailored to the human demonstrations of a specific player type through clustering. Training the affect model on a specific cluster (i.e., player type) will push the system to generate tracks that are tailored to their specific affective patterns. The specifics of the implementation of the evaluator are highly dependent on the environment and the type of content being generated, as discussed in Section \ref{sec:edrl_solid}. 

\subsection{Reward Function} \label{sec:reward}

The reward function is the final important component of our framework, which guides the designer to generate the desired stimuli. For example, if the generator is required to generate content that maximises arousal, it must be trained to produce stimuli that elicit $A_L$ close to the maximum arousal value. Beyond affect, the reward function can also be used to reward the similarity of state traces (i.e., in the behaviour of the evaluator). Generally speaking, the reward function, denoted by $R_L$, measures the similarity, $D$ between a generated affect trace ($A_L$) and the desired affect trace ($A_T$) as follows:

\begin{equation} \label{eq:reward}
    R_L = -  D(A_L, A_T)
\end{equation}

Similarity can be measured using any distance metric, $D$, considered appropriate for each use case. In this paper, we use the \emph{area between curves} method outlined in \cite{jekel2019similarity} as it is robust to noise and outliers. To reward for similarity, we then assign a negative value based on the distance (i.e., the greater the distance from $A_T$, the harsher the penalty) as seen in Eq. \ref{eq:reward}. It is important to note that the reward function proposed here can be used by either a traditional RL method (as a reward) or an evolutionary RL method (as a fitness function). 

\section{Arousal-Driven Racetrack Generation} \label{sec:testbed}

\begin{figure}[!tb]
    \centering
    \includegraphics[width=0.9\columnwidth]{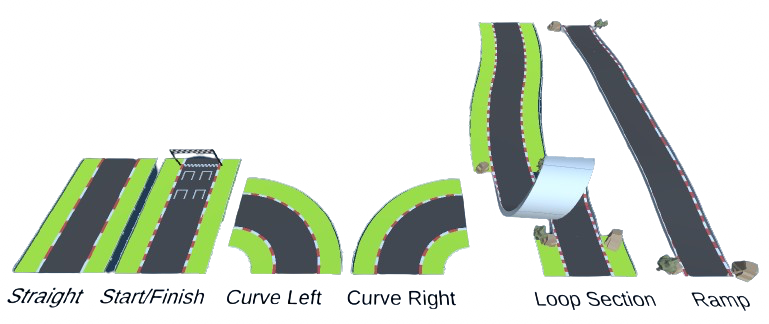}    
    \caption{Visualisation of the different possible track components found in \emph{Solid Rally}. }
    \label{fig:track_pieces}
        \vspace{-10pt}
\end{figure}

In this section, we briefly describe our case study platform, the \emph{Solid Rally} racing game from the AGAIN dataset \cite{melhart2022arousal}, along with an overview of our implementation (Section \ref{sec:edrl_solid}) and our arousal model (Section \ref{sec:arousal_model}). \emph{Solid Rally} is a 3D real-time rally driving game built on the Unity game engine. The player must race against three other opponent cars around a racing circuit featuring corners, straights, loops, and bridges (see Fig. \ref{fig:track_pieces}). As the cars drive around the racetrack, they are awarded points for passing through checkpoints located at the end of each component they drive through. The game features two sets of controls, one for steering (left, right, straight) and one for the gas pedal (forward, backward, neutral). The races have a maximum time limit of three laps, or 2 minutes if the player fails to finish before the time limit.

\subsection{Arousal Model for Solid Rally} \label{sec:arousal_model}

In this section, we describe the model of arousal used for evaluating the quality of the generated racetracks. As mentioned above, this system is built on top of \emph{Solid Rally}, a game from the AGAIN dataset \cite{melhart2022arousal} containing over 100 human demonstrations of continuous arousal traces. 

The observation data of the model consists of a set of 29 game features returned from the game engine about the cars' speed, score, distance to opponents, and various other in-game metrics. Each record in the dataset corresponds to a 3-second time window where the above features and arousal values are averaged. We then convert the records into preferences by comparing pairs of consecutive windows and giving them an ordinal affect label as denoted by the difference between their arousal values: increase, decrease, or stable arousal \cite{yannakakis2018ordinal}. Inspired by earlier studies \cite{barthet2022generative,makantasis2021pixels} we use a preference threshold of $0.15$, which means that for a time window to be labelled as an increase or a decrease, the corresponding absolute change must be greater than $0.15$. This threshold helps us combat reporting biases across subjects in the affect annotations provided. Finally, the dataset is normalised using min-max normalisation to ensure all the features lie within $[0, 1]$. 

\begin{figure}[!tb]
    \centering
    \includegraphics[width=0.9\columnwidth]{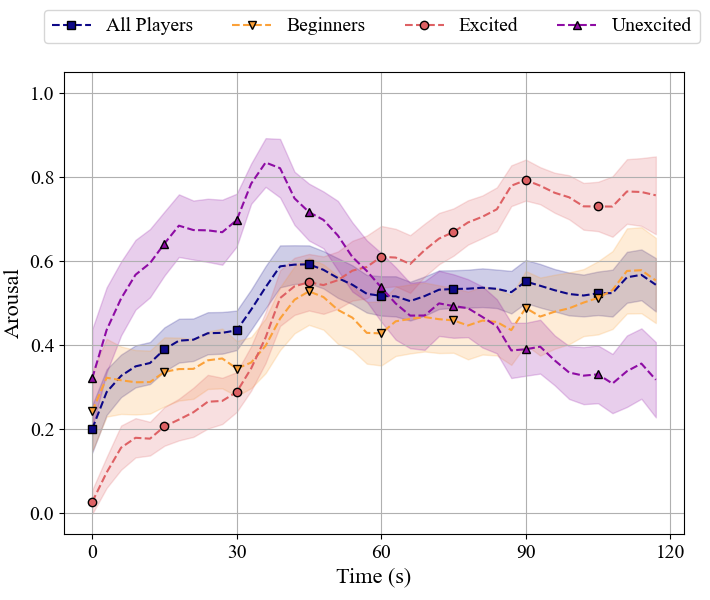}
    \caption{Mean arousal traces from all players and the three identified clusters of players within \emph{Solid Rally}, which we call the \textit{Beginners}, the \textit{Excited Experts} and \textit{Unexcited Experts}. Shaded areas denote a 95\% confidence interval.}
    \label{fig:cluster_arousal}
    \vspace{-10pt}
\end{figure}

To better differentiate between different annotator types, we cluster the dataset based on the player's score and arousal traces. We perform clustering by computing the distance between all possible pairs of annotators using the area between curves method described in Section \ref{sec:reward}. Doing so yields three clusters of annotators, two of which are a group of high-performing players with a substantial difference in their annotated arousal traces. We label these two clusters as \textit{Excited} experts and \textit{Unexcited} experts. Figure~\ref{fig:cluster_arousal} shows the mean arousal traces of these clusters, where we can clearly observe that \textit{Excited} experts have a constant increase in arousal over time, whereas \textit{unexcited} experts have a sharp rise at the start, followed by a gradual descent as players of this cluster get less and less aroused as gameplay progresses. The third cluster we identified hosts a group of poor-performing players in terms of behaviour, which we call \textit{Beginners}.

For the affect model, we implement a distance-weighted KNN model \cite{peterson2009k} to generate the change in arousal from a given pair of states, in this case, the evaluator's current state and the previous one. The KNN takes as input two state vectors (i.e., 29 game features across 3-second windows each as described above) and searches the corpus of affect annotations for the closest K entries using a pairwise Euclidean distance measure. We down-sample the dataset to only include the samples with changes in arousal (i.e., ignore stable entries) to focus on the most informative data samples during evaluation. Entries with an arousal increase are labelled with 1, whereas arousal decreases are labelled with 0. Once the K nearest neighbours are found, we use a weighted average (i.e., the closer to the current state, the higher the weight) to generate an arousal label for the current state. This label lies in $[0, 1]$ and can be interpreted as the change of arousal for the current state with respect to the previous state. The closer to 0 or 1, the more confident the agent is in its prediction of the change.

\subsection{Arousal-Driven PCG for Solid Rally} \label{sec:edrl_solid}

In this paper, we implement a generator using two designer approaches, the first is an evolutionary algorithm, often seen as an EDPCG approach, and the second is an EDRL agent using a variant of the Go-Explore algorithm \cite{ecoffet2021first} called Go-Blend \cite{barthet2022generative}. To accomplish this task, a framework had to be built around \emph{Solid Rally} to facilitate the creation of new racetracks, and communication between the internal game state and the generator's code. The game was converted into an Open-AI Gym environment using the Unity-Gym package, which allows for the training of Unity ML agents \cite{juliani2018unity} and custom agents using Python.

Racetracks are represented as a string of track components, the types of which are visualised in Fig. \ref{fig:track_pieces}. Each component has a start position and end position, which are used to connect the pieces during generation. The components are scaled to be the same unit size, i.e., the straight, start/finish, and curves are all 1 \textit{tile} long, whereas the loop and ramp are 3 \textit{tiles} long. This tile-based formatting allows us to represent the environment as a 2-dimensional grid of tiles, simplifying the process of detecting collisions and determining feasibility. The grid is initialised to contain zeros (i.e., empty tiles) at the start of training. 

\subsubsection{EDPCG Designer}

Our implementation of the EDPCG designer for this case study is as follows. The genotype is a string of IDs of a fixed target length, which are mapped to the components and instantiated in sequence in the environment. The phenotype is generated by iteratively replacing the component ID found in the genome with the corresponding track piece and updating the 2D grid accordingly. Collisions are detected by checking whether the current location already has a component (i.e., an ID  $\neq \emptyset$) before placing the next piece. If the racetrack is found to collide with itself, the stimulus is considered infeasible and is assigned a high penalty value (i.e., the worst possible reward of $-1000$). 

We employ a genetic algorithm to allow the EDPCG designer to search the solution space for highly fit tracks. Our implementation follows the ($\mu-\lambda$) evolutionary strategy \cite{back1993overview}, with the addition of a crossover operation during reproduction. This works by first initialising a population consisting of random individuals. Then, across many generations, we pass each individual to the evaluator if they are considered feasible, and assign them a fitness as the reward $R_L$ (see Eq. \eqref{eq:reward}) based on their generated $A_L$. After the entire population is evaluated, we select the $\mu$ best individuals to act as parents for the next generation. The next generation of individuals is created by repeatedly selecting pairs of parents and using one-point crossover and mutation to generate offspring. We use uniform mutation, meaning that we randomly change components of an individual to a different ID with a small chance set by the mutation rate. We preserve the best individual between generations (elitism) to ensure the population retains feasible and high-quality individuals after reproduction. Once this process is completed (i.e., the maximum number of generations is reached), the designer outputs a set of high-quality tracks that should elicit the desired affect pattern.

The hyperparameters for the EDPCG designer in this use case are selected based on preliminary experiments and are as follows. We use a parent population size ($\mu$) of the best 10 individuals from the current population. Our offspring population size ($\lambda$) was set to 50 individuals, with the constraint that each individual in the population must be unique (i.e., when a new individual is generated, it is only inserted into the population if an identical copy does not already exist). We use one-point cross-over and mutation, with a mutation rate of 10\% during reproduction. Parents are randomly chosen for reproduction from the $\mu$ best individuals. We ran 10 separate runs of 50 generations per scenario described above, and present the average results across runs, including the 95\% confidence interval. 

\subsubsection{EDRL Designer}

As an alternative to the search-based generator, we use Go-Explore as the EDRL designer algorithm for this experiment, as it has proven itself a strong candidate for exploring deceptive and challenging environments. This designer follows the same collision rules as the EDPCG designer, but is driven by a different optimisation process. The EDRL designer iteratively builds racetracks using the Go-Explore exploration loop. First, a cell is sampled from the archive and its state is produced in the environment by repeating its trajectory of actions. From there it explores a single action, assigns a reward ($R_L$; see Eq. \eqref{eq:reward}), and checks whether to store it in the archive or not. Cells are stored if they are unique, or if they have a larger reward than the existing cell. Our cells' state representation is a state vector containing the number of straights, turns, loops, bridges, and the length of the Dijkstra path \cite{dijkstra}; described in the following section.
This more abstract representation was chosen to group similar racetracks into the same cell and only separate meaningfully different layouts, resulting in a smaller archive and better search space exploration. Finally, we cap the cells' trajectory length to a maximum of 10 pieces to maintain a fair comparison with the EDPCG and Random designers, meaning that Go-Explore cannot explore a cell if it already contains 10 actions. Once complete, we take the best cell from the archive with 10 actions as our elite individual.

\subsubsection{Playable Racetracks via Dijkstra}

Since the tiles placed by the designer are not guaranteed to create a racetrack that forms a circuit (i.e., a connected start and finish tile), the PCG algorithms are required to find the shortest path between the first and last tile, connect them, and form a playable track. With this in mind, we use the well-established Dijkstra algorithm \cite{dijkstra} to search the grid for the shortest path between the start and the end tile. Once the shortest path is found, Dijkstra's algorithm places appropriate tiles to fill the path and complete the circuit. Dijkstra is only allowed to use simple tiles (i.e., straight, curve left, curve right) to implicitly encourage the designer to not rely heavily on this method and, preferably, close the circuits through its own designs. In case no path can be found using Dijkstra---due to the absence of a feasible route between the start and the end tile---the track is marked as infeasible and is discarded.

\subsubsection{Evaluator}

The evaluator consists of an in-game agent that drives around the generated racetracks and queries the arousal affect model described in Section \ref{sec:arousal_model}. The agent's behaviour is governed by a checkpoint system that forms part of the original \emph{Solid Rally} implementation, which works as follows. Each component of the generated level has a checkpoint placed at its endpoint. The agent drives such that it minimises the distance to the next checkpoint. When the agent drives through this checkpoint, it is given an increment to its score, and its target is set to the next checkpoint. This simple system allows the agent to behave reliably across any kind of level generated. This is the same system used by the opponent cars which the evaluator is racing against and---like the original game---the race takes place over 3 laps. As the evaluator simulates a race, it queries the KNN affect model every 3 seconds of in-game time. Note that due to the relatively deterministic nature of the evaluations, we only perform one simulation per track; however, in less deterministic game environments averaging the $R_L$ across multiple independent runs would be necessary for yielding more stable training signals.

\section{Experimental Protocol} \label{sec:protocol}

We evaluate generators by testing them across three different scenarios by which they must generate content that elicits a desired (i.e., target) affect trace to a game-playing agent. As mentioned in Section \ref{sec:reward}, we use the area between curves as the distance measure between the generated affect trace ($A_L$) and the target trace ($A_T$). The content was tested by simulating three laps around the racetrack to remain consistent with the original dataset. The first scenario is the \emph{Minimise Arousal} experiment, where the generator must create a racetrack that causes the agent's arousal to constantly decrease over time. The \emph{Maximise Arousal} scenario mirrors the former by requiring the agent's arousal to constantly increase over time. Finally, the \emph{Fluctuating Arousal} requires the racetrack to elicit an arousal trace that varies over time. In particular, the level must first maximise arousal for its first third, then minimise arousal for the second third, and finally, for the last third of the track, maximise arousal for the player. We compare our affect-driven level designers against a baseline random designer agent, which places a series of random tiles and then uses Dijkstra to close the circuit. 

To evaluate the generators, we pick the best individual based on the \textit{accuracy} of the output arousal signal elicited from the generated track to the target signal. Specifically, we treat the arousal modelling of the generated track as a binary classification problem, and we measure the rate at which the output signal and the target signal agree on the change in affect (i.e., increase versus decrease). Since our arousal model outputs a value between 0 and 1, an increase (or decrease) is indicated by any output over (below) $0.5$. An accuracy of $100\%$ (or $0\%$) means that the generated arousal trace and the desired arousal trace agree (disagree) completely on the change of arousal across the complete racetrack. Comparisons are made across 10 runs for each experiment configuration. We report average accuracy values and corresponding 95\% confidence intervals. We denote significance when we observe non-overlapping confidence intervals between experiments at the $p < 0.05$ level. Beyond this, we also perform an expressive range analysis on the output of the designers \cite{smith2010analyzing}. The expressive range analysis offers us complementary qualitative insights about the variation of the components used on the racetrack for each target signal.

\begin{table}
   
    \caption{Accuracy of the best individual from each of the three scenarios (minimise, maximise or fluctuate arousal), grouped by designer algorithm (random, EDPCG, EDRL) and averaged across 10 runs, including 95\% confidence intervals. Values in bold indicate the highest accuracy value obtained across designers for a particular scenario. Underlined and bold values indicate statistically significant differences across designers.}
    \label{tab:results}
    \renewcommand{\arraystretch}{1.35} 
    \begin{adjustbox}{width=\columnwidth}
    \begin{tabular}{|c|c|c|c|c|} 
         \hline
         {\textbf{}} & \textbf{All Players} & \textbf{Beginners} & \textbf{Excited} & \textbf{Unexcited}\\ 
         \hline
         \hline
         \rowcolor{gray!30} 
         \textbf{Random}& &    &  &  \\ 
            \hline
            Max. Arousal & $72.4\pm2.8$ & $58.0\pm2.6$ & $50.7\pm3.5$ & $89.8\pm1.0$ \\
            \hline
            Min. Arousal & $66.4\pm2.5$ & $76.9\pm1.7$ & $89.2\pm1.9$ & $35.8\pm3.6$ \\
            \hline
            Fluctuating & $57.1\pm1.7$ & $72.3\pm3.6$ & $56.9\pm1.6$ & $63.5\pm2.0$ \\
            \hline
         \hline
         \rowcolor{gray!30} 
         \textbf{EDPCG}& & &  & \\ 
            \hline
            Max. Arousal & $84.4\pm2.0$ & $\bm{78.8\pm2.0}$ & $70.4\pm7.7$ & $95.9\pm0.9$\\
            \hline
            Min. Arousal & $\bm{81.6\pm2.2}$ & $85.8\pm2.1$ & $96.6\pm1.0$ & $42.8\pm2.6$  \\
            \hline
            Fluctuating & $68.3\pm2.2$ & $86.3\pm1.0$ & $67.2\pm3.0$ & $71.1\pm1.6$ \\
            \hline
         \hline
         \rowcolor{gray!30} 
         \textbf{EDRL}& & & & \\ 
                  \hline
            Max. Arousal & $\bm{87.7\pm2.8}$ & $\bm{78.8\pm3.4}$ & $\bm{75.9\pm3.9}$ & $\bm{97.4\pm0.8}$\\
            \hline
            Min. Arousal & $\bm{81.6\pm2.9}$ & $\underline{\bm{96.2\pm1.5}}$ & $\bm{98.1\pm1.2}$ & $\underline{\bm{53.4\pm2.8}}$\\
            \hline
            Fluctuating & $\underline{\bm{74.7\pm2.5}}$ & $\bm{87.5\pm2.4}$ & $\underline{\bm{75.3\pm2.1}}$ & $\bm{72.7\pm0.9}$\\
            \hline
    \end{tabular}
    
    \end{adjustbox}
    \vspace{-10pt}
\end{table}

\section{Results}

The results of our experiments comparing the three different track designers can be seen in Table \ref{tab:results}: we test each designer across four different player clusters and three target arousal scenarios. As expected, we observe that both the EDPCG and EDRL designers outperform the random designer (significantly based on the 95\% confidence intervals) across all player clusters and scenarios. It is also clear from the results that certain scenarios were more challenging than others across player clusters. For example, the \textit{excited} and \textit{unexcited experts} produce the most varied results across the three different scenarios tested. \textit{Unexcited experts} are easier to satisfy as a player group when we attempt to maximise their arousal with accuracy values over 95\% for both EDPCG and EDRL. The opposite holds for the \textit{excited experts} and the \textit{beginners}, as it seems that minimising arousal is easier than maximising arousal for these player groups. Experiments with \textit{all players} available in our corpus yields overall more consistent results across designers and scenarios (i.e., accuracy values in between 68\% and 87\%) due to the larger and more diverse set of human demonstrations. This larger and more diverse corpus most likely allows the designer to explore a larger variety of arousal states and therefore yield more consistent performances.

We argue that the varying performance of the tested methods across player types is due to their dissimilar affective patterns (see Fig. \ref{fig:cluster_arousal}). We argue that the inherent subjectivity of reported affect as elicited by the layout of the circuit causes such dissimilarities. For example, compared to other player types, the arousal changes of the \textit{beginner} players appear to be more consistent to the design of the circuit (i.e., the context). Consistent affect demonstrations, in turn, make the task of content generation far easier as predicting any target arousal trace is much simpler. On the other end of the spectrum, the \textit{unexcited experts} generally showcase less arousal variation with respect to racetrack changes. We argue that this is due to habituation effects, as this group of expert players appears to be stimulated less after driving their first lap. Another potential reason for the observed differences is that the skill level of the player clusters had a strong effect on the annotated arousal levels. Specifically, our evaluator agent showcases only marginally better performance to the opponent AI cars, which is roughly equivalent to the behaviour of the \textit{beginners} cluster. Such an agent cannot replicate how either of the expert groups would play in any generated racetrack, and it thus results in skewed arousal levels and larger performance discrepancies across scenarios for these groups.

As seen in Table \ref{tab:results}, EDRL yields superior performances compared to the random baseline while it matches or outperforms the performance of EDPCG across nearly all scenarios and player types. Results show that EDRL is, on average, able to significantly outperform EDPCG in 4 out of the 12 scenarios explored and yields higher performances in 10 out of those scenarios. Intuitively, we argue that the EDRL designer is able to better explore the search space over time, whilst the EDPCG designer is more prone to converging to local optima through the simple evolutionary strategy implemented. 

Finally, we also analyse the differences in track layouts across the scenarios we tested. Unsurprisingly, tracks built to minimise arousal appear to contain significantly more basic tiles (i.e., straights and turns) compared to tracks built to maximise arousal, with averages of $24.3\pm1.6$ and $19.7\pm2.5$ respectively. We observe the inverse phenomenon when we attempt to maximise arousal as generators use more complex event tiles (i.e., bridges and loops) to induce increasing arousal whenever possible; an example of this phenomenon can be seen in Fig. \ref{fig:solid_rally}. Placing more event tiles in the maximise arousal scenario---on average $5.1\pm0.9$ tiles compared to $3.8\pm0.5$ tiles in the minimise arousal scenario---increases the probability of driving errors (e.g., driving off-road, crashing into other cars, barriers) from our evaluator agent. Such driving style can, in turn, contribute to increases in arousal elicitation. 

\section{Discussion}

The proposed approach is novel in that it generates affect-aware content in a continuous manner via the EDRL framework. Importantly, the generated content can be tailored to a target affective pattern or trace for a particular user. Our results show that EDRL is capable of matching and sometimes outperforming EDPCG when it comes to designing racetracks with a specific arousal trace as a target. Incorporating Go-Explore's \cite{ecoffet2021first} robustification phase would allow our EDRL agent to function in an online capacity. This superior performance paired with the ability of EDRL to act as an online content generator (e.g., as in \cite{wang2022fun,Experience-Driven-PCG}) highlight the promise of this approach in a real-world setting. Future investigations employing more complex reward functions, tested across dissimilar game genres and other domains, will help us validate the potential of EDRL for affect-driven generation in a general fashion. While tested initially in the domain of games, the EDRL method introduced here is applicable to any affective interaction task that envisions closing the affective loop via content generation mechanisms. As EDRL has shown promise to adapt itself to clusters of users and target affect patterns in this paper, we envision methods relying on the EDRL principle to be able to cope with dynamic shifts of user preferences and behaviours over time.

One of the limitations of the current approach lies in the simple behaviour of the evaluator agent. Whilst we chose the checkpoint system that governs its behaviour---due to its reliability to generalise to new levels---the resulting behaviour of the agent is very deterministic and does not follow human-like patterns of play. More specifically, the agent does not factor in opponent cars, so it is prone to collide with them in situations that most human players would likely avoid. The agent is also constantly accelerating (a behaviour that most human players would likely not follow) especially if they get stuck or need to slow down for a sharp turn. Consequently, since the agent tends to play through components with very similar behaviour, it limits the generator's ability to thoroughly explore the solution space as the full range of possible scenarios per component is not sufficiently explored. One way of addressing this limitation in the future is to train agents that behave similarly to the target clusters, thus having a more representative evaluator in terms of behaviour and, therefore, more reliable affect-driven generation.

Another limitation of this work lies in the generalisability of our arousal model. Since the dataset used in this paper only contains annotations and telemetry from sessions on a single-track layout, the arousal model does not have a large distribution of contextual data available, which in turn limits its performance when assessing new content. As a result, whilst we show that EDRL is capable of generating content using this arousal model, it remains to be seen whether these results will accurately reflect human feedback. This is an important avenue for future work, as quantitatively correlating the generated tracks via a human study (e.g., in \cite{yannakakis2009real}) will confirm that the generator is capable of matching content to human emotions.

Finally, collecting more human (behavioural and affect) demonstrations on a diverse set of new racetracks will help us build more general representations between arousal, playing behaviour and game content. Yet another approach could be a mixed initiative one \cite{yannakakis2014mixed}, where a human user provides feedback on the generated tracks to help steer affect-driven optimisation in the right direction. This approach could gradually help us build a more personalised model of affect, and---with enough data obtained---could remove the need for running simulations and instead learn to generate affect directly from the stimuli presented. 

\section{Conclusions}

We introduced a novel framework that expands current experience-driven content generation frameworks \cite{EDPCGRL,Experience-Driven-PCG} able to generate content that elicits tailor-made continuous affective patterns to a particular type of user. We test the capacity of two such generative methods---one based on a genetic algorithm and a second one driven by RL exploration---to create affect-aware and personalised racetrack levels in games. Our results demonstrate that both approaches can accurately match a number of desired affect traces for certain types of players, but the EDRL method appears to be more efficient and robust. Both methods, however, face challenges to elicit certain affect patterns (e.g., maximise arousal throughout the level) to player types that have not experienced such patterns through their demonstrations (e.g., low-aroused beginner players). Findings of this paper suggest that it is possible to continuously generate affect-driven content and, importantly, it appears that the proposed EDRL method can create personalised and tailor-made content for various user types. While the results of this paper are specific to game content generation, the methods proposed are directly applicable to any affective interaction domain in need of personalised content creation.

\section*{Ethical Impact Statement}

This paper makes use of an existing dataset (AGAIN dataset \cite{melhart2022arousal}) of human demonstrations collected from crowd workers on the Amazon Mechanical Turk (mTurk) platform. This dataset is publicly available, and participants gave their consent for their data to be stored and utilised in an anonymous fashion. To the best of our knowledge, there is no significantly negative application of the methods we use in this paper and no added privacy or discrimination risk. The data used in the paper and environment are publicly available for scientific reproducibility and for further extensions of this study.

\bibliography{bibliography}
\bibliographystyle{ieeetr}

\end{document}